\documentclass[%
 aip,
 amsmath,amssymb,
 reprint,%
]{revtex4-1}

\usepackage{graphicx}% Include figure files
\usepackage{dcolumn}% Align table columns on decimal point
\usepackage{bm}% bold math
\usepackage[utf8]{inputenc}
\usepackage[T1]{fontenc}
\usepackage{mathptmx}
\usepackage{etoolbox}

\usepackage{xcolor}

\makeatletter
\def\@email#1#2{%
 \endgroup
 \patchcmd{\titleblock@produce}
  {\frontmatter@RRAPformat}
  {\frontmatter@RRAPformat{\produce@RRAP{*#1\href{mailto:#2}{#2}}}\frontmatter@RRAPformat}
  {}{}
}%
\makeatother
\begin{document}

\preprint{AIP/123-QED}

\title[Loading of a large Yb MOT on the $^1S_0 \rightarrow ^1P_1$ transition]{Loading of a large Yb MOT on the $^1\textbf{S}_0 \rightarrow ^1\textbf{P}_1$ transition}
\author{Hector Letellier}
\author{Álvaro Mitchell Galvão de Melo}
\author{Anaïs Dorne}
\author{Robin Kaiser}%
 \email{robin.kaiser@univ-cotedazur.fr.}
\affiliation{Université Côte d’Azur, CNRS, INPHYNI, UMR7010, 17 Rue Julien Lauprêtre, 06200 Nice, France}
\date{\today}% It is always \today, today,

\begin{abstract}
We present an experimental setup to laser cool and trap a large number of Ytterbium atoms. Our design uses an oven with an array of microtubes for efficient collimation of the atomic beam and we implement a magneto-optical trap of $^{174}Yb$ on the $^1S_0 \rightarrow ^1P_1$ transition at $399 nm$. Despite the absence of a Zeeman slower, we are able to trap up to $N=10^9$ atoms. We precisely characterize our atomic beam, the loading rate of the magneto-optical trap and several loss mechanisms relevant for trapping a large number of atoms. 
\end{abstract}

\maketitle

\section{\label{intro}Introduction}

Laser cooling and trapping of atoms has been a very active field of research since the 80s \cite{Cohen-Tannoudji1998, Chu1998, Phillips1998}, with a large variety of elements trapped and cooled and with a wide range of physical phenomena studied. Initially, the choice of elements used to study laser cooling and trapping has been a compromise between elements having closed two level schemes and the availability of convenient laser sources. This explains the initial dominance of sodium using dye lasers at $589 nm$ and, with the development of semiconductor lasers for large scale commercial applications, rubidium and cesium atoms in the early 90s (with some exceptions, such as metastable helium when specific laser expertise was available in the laboratories\cite{Vansteenkiste1991}). After the realization of Bose-Einstein condensation \cite{Cornell2002, Ketterle2002} and the rapid development of readily available laser sources over a large range of frequencies\cite{Hollberg1991}, many more species have been laser cooled and trapped, allowing the research community to target novel phenomena to be studied when choosing appropriate elements. Alkaline earth atoms share a particular property distinct from most other elements by having bosonic isotopes with a non degenerate (spin zero) ground state. In this case, Sisyphus cooling mechanisms do not apply as they do e..g in alkali atoms. Other elements share this property of a non-degenerate (spin zero) ground state where only Doppler cooling and trapping are at work. This paper presents our new experimental scheme on laser cooling and trapping of Ytterbium atoms \cite{Nomura, YbCsTrap, CrossBeam_Plotkin, ThermalSource, Guttridge_2016, QuantumYbgas, Wodey_2021, CoreShellMOT, Kuwamoto, Kawasaki_2015, MultiIsotopesTrap, Zhao_Peng-Yi_2008}, with several bosonic isotopes with a zero spin ground state. The choice of Ytterbium atoms (in comparison e.g. with Mg \cite{Sengstock1994, Loo2004}, Ca \cite{Binnewies2001, Curtis2001} or Sr \cite{Katori1999, Vogel1999, Bidel2002, Chaneliere2005, Yang2015} atoms) is a convenient set of parameters to study Anderson localisation of light in three dimensions\cite{Skipetrov2014, Bellando2014, Skipetrov2016, celardo2018}. This goal requires, in particular, the trapping and cooling of a very large sample of atoms and a large optical depth, with only a moderate requirement of the lifetime of trapped atoms, as a few ms is sufficient to study multiple scattering of light. In the present paper, we present our experiment setup aiming at a compact system allowing us to trap a large number of atoms. 
Section \ref{Experimentalsetup} presents the vacuum setup, the Ytterbium oven, the laser at $399 nm$ and transverse laser spectroscopy on the  $^1S_0 \rightarrow ^3P_1$ line. In section \ref{MOTModel} we recall the model for  loading of atoms in the single particle regime as well as when including light-assisted collisional losses. Our experimental results are presented in section \ref{Experimentalresults} and a conclusion in section \ref{Conclusion}

\section{\label{Experimentalsetup}Experimental setup}
\subsection{\label{Design}Vacuum setup, oven and nozzle}

\begin{figure}
\centering
\includegraphics[width=0.49\textwidth]{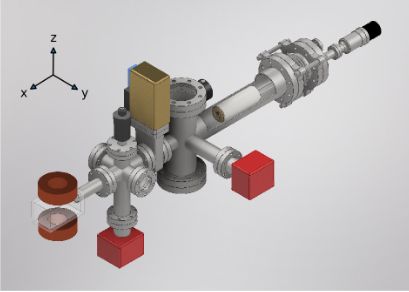}
\caption{\label{fig:planvide} Scheme of the experimental system, with the oven and bellow on the right, two CF63 window ports, a pneumatic valve, and two CF35 window ports. In red, two ion-getter pumps and on the left, the glass science cell and the coils for the magnetic field gradient (with the strongest field gradient along the vertical z-axis). On the first UHV cross before the pneumatic valve, one port is used for pre-pumping with a manual UHV valve. The second UHV cross (after the pneumatic valve) also contains a mechanical blocker, discussed at the end of the paper.}
\end{figure}

The design chosen for our experiment is shown in Figure \ref{fig:planvide}. One important point to notice is the absence of a Zeeman slower. This choice is motivated by the more compact system, in addition to the absence of any stray magnetic field to be managed at the position of the magneto-optical trap (MOT). 
We use an oven to heat up a sample of metallic Ytterbium (Yb) with a natural abundance. The oven is based on a commercial effusive source (OEM S40 DZ35 K, Riber), used for evaporation of various metals up to well above 1000°C. 
For our experiments, we run the oven at much lower temperatures, in the range of $300-500^{\circ}$C. 

This oven comes along with water cooling inside the vacuum chamber, allowing to strongly reduce the outer temperature of the vacuum chamber with correspondingly reduced thermal nuisance on the optical table. 
This oven contains two heating filaments, allowing to heat separately a crucible body and output nozzle. A  shutter, mechanically accessible from the outside, is also included in front of the oven. A small tipping of the oven is corrected using a bellow connecting the oven to the vacuum chamber to ensure the horizontality of the setup.

\begin{figure}
\centering
\includegraphics[width=0.49\textwidth]{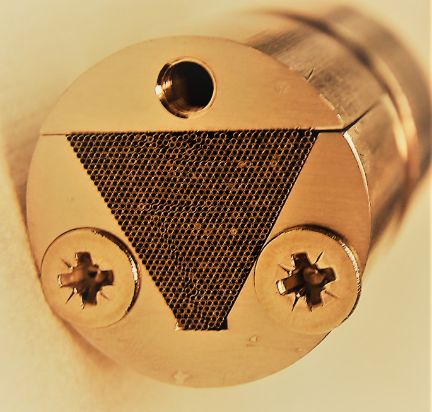}
\caption{\label{fig:capi} Image of the output facet of the nozzle of our atomic oven: an array of well aligned micro-tubes of length $l_c=13$ mm, $d_i=250 \mu m$ inner and $d_o=340 \mu m$ outer diameter respectively is used to provide a collimated beam of atoms.}
\end{figure}

In order to obtain a well collimated atomic beam, we use an array of micro-tubes (see Figure \ref{fig:capi}). 
Such a micro-tube arrangement increases the lifetime of the oven for a given on-axis flux of atoms. We have used  stainless steel (AISI-304) micro-tubes of length $l_c=13$mm, $d_i=250 \mu m$ inner, and $d_o=340 \mu m$ outer  diameter respectively. The corresponding geometrical aperture is thus $\theta_c = 19$mrad. Each micro-tube thus produces a well collimated beam. It is however also necessary to secure that all micro-tubes are well aligned, as random orientation of them would produce a effective broadening of the atomic flux. 
A regular ordered stacking of tubes produces a hexagonal pattern. Such a pattern is a good signature of the alignment between the micro-tubes. In order to allow this hexagonal pattern to be compatible with the boundary conditions of the support structure we choose an equilateral triangular structure \cite{Senaratne}. For mechanical reasons, the bottom apex has been flattened, without breaking the hexagonal pattern. This nozzle has been inserted in a custom made ceramic crucible, allowing for better heat management than a stainless steel crucible. 
This periodic structure shown in Figure \ref{fig:capi} is an excellent indicator of the alignment of our micro-tubes. We have also verified that beyond the mere good transmission of an incoherent light source through our array, which could also be achieved by channeling of light through a misaligned arrangement of micro-tubes, we were able to identify a structured picture by looking through our array. 
We note that the use of such a micro-tube array does not increase the on-axis flux of atoms at a given oven temperature. It however strongly reduces the off-axis atomic flux (at least in the molecular regime discussed below) and thus increases the lifetime of the oven with a given load of Yb by an estimated factor of 40 in our case and it also reduces the required pumping speed of the vacuum pump in the oven chamber.

We have used a ultra-high vacuum (UHV) cross to connect a first ion getter pump (75l/s-N2, NEXTorr Z200, SAES) opposite to a manual UHV valve (Mini UHV Gate Valve CF40, VAT) and two window ports allowing for transverse spectroscopy of the atomic beam at the output of the oven (and if necessary in future, transverse cooling and collimation of the atomic beam). The output face of the nozzle is placed 8cm before the transverse section of the UHV cross, preventing thus deposition of the windows by Ytterbium from the oven.

The oven part of the vacuum chamber is separated from the science cell by a pneumatic valve (Mini UHV Gate Valve CF40, VAT), preserving the vacuum of the science cell during the initial heating of the oven and when a refill of metallic Yb is needed. We also added two differential pumping tubes (length=40 mm, diameter=12 mm) before and after the pneumatic valve. Between the pneumatic valve and the science cell, we have added a second UHV cross, required to connect a second ion getter pump, identical to the first one, allowing for pressure below $10^{-9}$mbar in the science cell with a  $10^{-8}$mbar at the output of the oven during its operation.  The cross also features two window ports for further spectroscopy and one mechanical shutter (whose goal will be discussed in this paper).
The science cell, where we trap atoms and perform all experiments with cold atoms, is of rectangular shape (50mmx50mmx100mm), with broadband (399nm-700nm) coating on each interface of the borofloat windows (Japan Cell). The use of a glass science chamber also allows for convenient and flexible optical access and strongly reduces any Eddy currents when switching off the magnetic field gradient (using two coils in anti-Helmholtz configuration, positioned above and below the science cell, see Figure \ref{fig:planvide}) required for MOTs before performing experiments in free fall. 

The distance between the nozzle output and the center of the science cell where atoms are trapped is 52cm, the cell entrance limiting the angular aperture to 50mrad. This is larger than usual experiments using Zeeman slowers, and contributes to the compact and robust operation of our experiment. We note that our design, like most others, suffer from the deposition of Yb atoms at the cell window facing the oven. This prevents the efficient use of an on-axis slowing beam without resorting to optical cleaning methods of the cell inner surface. We note that in contrast to similar situations using Strontium atoms, Ytterbium is a poor metal (with , preventing its use as a mirror (with e.g. an angle at 45° placed inside the science cell). 

\subsection{\label{LaserSpectro}Laser system and spectroscopy of the atomic beam}

\subsubsection{\label{Lasersystem}Laser at 399nm}

\begin{figure}
\includegraphics[width=0.45\textwidth]{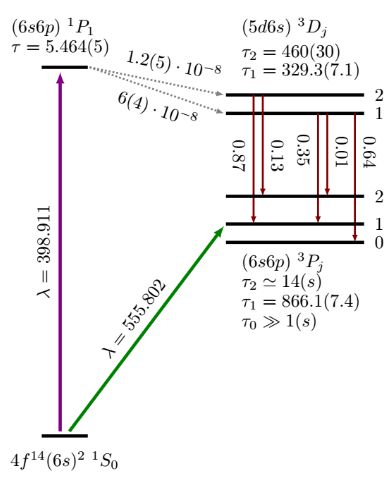}
\caption{\label{fig:Yb} Energy level diagram showing the relevant transitions and branching ratios for our experiment. All lifetimes are given in $ns$, except if otherwise specified.}
\end{figure}

Laser cooling and trapping of Yb atoms are possible due to closed transitions, both at 399nm and 556nm (see Figure \ref{fig:Yb}). As the transition at 399nm is dipole allowed with a natural linewidth of $\Gamma/2\pi=29 MHz$, the capture range using this atomic transition is much larger than the one at 556nm with a linewidth of $2\pi 182kHz$. 
We therefore choose to implement a MOT at 399nm, even though we will use the 556nm line for further cooling in the future. As for our research project, trapping a large number of atoms is an important feature, we use a high-power frequency-doubled TiSa laser. Indeed, as we will discuss below, the number of trapped atoms increases rapidly with the size of the trapping laser, provided one can secure a large saturation intensity in such large laser beams. In many experiments with Yb, the available power on $^1S_0 \rightarrow ^1P_1$ transition at $399nm$ with a saturation intensity of $I_{sat} = 60mW/cm^2$, was a strong limitation, requiring the use of smaller beam diameters.
In our experiment, we use a Ti:Sapphire laser (M2 SolsTiS) pumped by an 18W frequency-doubled ND:YAG laser (Verdi18, Coherent), providing 6W at 798nm. We frequency double this laser in a resonantly enhanced frequency-doubling cavity (M2 ECD-X). We can thus generate up to 2.5W of laser light at 399nm. We lock the frequency of the  laser (at 798nm) using a feedback loop to a wavemeter (High Finesse WS8-2) with a 2 MHz precision, sufficient for the transition at 399nm with a linewidth of $\Gamma = 2\pi 29MHz$. 
In order to realize a stable setup with minimal spatial inhomogeneities, we inject the output of the laser in an optical fiber (crystallographic UV prototype fiber, NKT, LMA-PM-10-UV with wet silica as a raw material) at the entrance of a compact free-space commercial optical bench (Muquans) with glued optical components and acousto-optic modulators (AOM) (MQ110-A3-UV, AA Optics). This bench delivers the 6 output ports for the MOT (controlled by the same AOM and an additional mechanical shutter, and also with individual power control using $\lambda/2$ plates) and one additional probe beam with its own AOM and shutter. 
We also use a slowing beam counter-propagating to the atomic beam. This slowing beam (with less critical beam and polarisation specifications) is directly separated after the frequency doubling cavity. This beam is passed through an AOM (Gooch$\&$Housego I-M110-3C10BB-3-GH27) with a diffraction efficiency of $92\%$. This slowing beam has a power of 300mW and is red-detuned from the MOT laser beams by 220MHz. Note that this frequency difference is fixed, as both the slowing beam and the MOT beams are derived from a single laser whose frequency is set by the frequency of the Ti:Sapphire laser.

The optics of the 6 independent MOT laser beams at the science cell is based on an optical cage system (Thorlabs) holding the polarisation control elements ($\lambda/2$ and $\lambda/4$ waveplates) and a final beam expander (a f=7.5mm aspheric lens followed by a 2" f=150mm plano-convex lens with a final beam waist of $w_0=22mm$).
When using the commercial optical bench for the beam splitter, we use a total power of $P_0=100mW$ in the 6 beams at the MOT position. This corresponds to a center beam intensity of  $I = 2P_0/6\pi w_0^2 \simeq 0.037I_{sat}$. Using higher input power in the fiber to the division bench allows for larger saturation, but in order to avoid damaging the fiber, we have so far restricted the injection to a maximum of 300mW. Some experimental results presented in this paper have made use of larger beam power, using an independent  homemade free space setup. We stress that the proper intensity balance as well as a homogeneous laser beam are the key to stable MOT operation at large atom number, in particular at low magnetic field gradients, arguing in favor of the use of the commercial beam divider with the spatial filtering provided by the optical fiber.

%Thus the intensity at the center of each beam is $I = 2P_0/6\pi w_0^2 \simeq 0.08I_{sat}$. The capture velocity of this molasses is $kv_c/\Gamma = 3 \rightarrow v_c = 34m/s$. With the same parameter the MOT capture velocity is $42m/s$. Not so different and optimum gradient $ $.

\subsubsection{\label{AtomicBeam}Atomic beam flux and spectroscopy}

Efficient loading of a large number of cold Ytterbium atoms from an atomic beam requires a quantitative knowledge of the atomic flux arriving in the science cell. 
We thus first estimate the atomic flux exiting the oven as a function of its temperature $T$. Assuming a molecular flow regime, i.e. the atom mean free path $\bar{\ell}$ is much larger than the diameter of the micro-tubes, the conductivity of the aperture is given by the product of its area and the mean atomic velocity: 
\begin{eqnarray}
    C_{mol} = A \bar{c}/4 = N \cdot 3.5 \cdot 10^{-3}\ l/s
\end{eqnarray}
where $A=N\pi d_i^2/4\simeq 5N \cdot 10^{-8} m^2$ is the open area of $N$ micro-tubes, $\bar{c}$ the mean thermal velocity ($\bar{c}=\sqrt{8 k_B T/\pi M}$ and $M$ the mass of the $^{174}Yb$ atoms). The conductivity of our micro-tube array is the product of the above aperture conductivity and the transmission factor. For long micro-tubes ($d_i \ll l_c$) the transmission or Clausing factor \cite{Clausing} is $t_{Cl} = 4d_i/3l_c = 2.6\%$.
We note that by neglecting here the open area between the micro-tubes, which contributes 10\% additional surface, but with a narrower transverse size, the total transmitted flux is only increased by 2$\%$. Finally, to estimate the output flux of the nozzle one still needs to compute the atomic density inside the oven ($n=p/k_B T$), where $p$ (in Pa) is the total saturated vapor pressure of Ytterbium as a function of the temperature T (in K):
\begin{eqnarray}
    \log(p) &=& 14.117 - 8111/T - 1.0849\ log(T).
    \label{EqPressure}
\end{eqnarray}
The total flux of Yb atoms leaving one micro-tube is thus given by:
\begin{eqnarray}
      \varphi_{c1} &=& n\ C_{mol}\ t_{Cl},  
\end{eqnarray}
which at $T = 400^{\circ}$C is of the order of $\varphi_{c1} \simeq 1 \cdot 10^{12} at/s$. The total flux of atoms leaving the oven with $\sim 700$ micro-tubes is thus of the order of $\varphi_c \simeq 7 \cdot 10^{14} at/s$.
Note that one needs to take into account the isotopic abundance to estimate the flux of a specific isotope (corresponding e.g. to $32\%$ for $^{174}Yb$).

In order to assess in which regime our atomic oven is operating, we use the ratio between the mean free path and the spatial dimensions of the micro-tube, which is the so-called Knudsen number $K_n=\ell/d_i$. For $K_n \gg 1$ the atomic flow is in the molecular regime, with rare inter-particle collisions and atoms mainly interacting with the walls of the micro-tubes. When increasing the temperature of the oven in order to increase the atomic flux, the mean free path decreases and the flow stops to be in the molecular regime and enter a laminar and finally a turbulent flow regime. 
The mean free path $\bar{\ell}$ depends on the temperature $T$ of the gas, its pressure $p$ and an inter-particle interaction characterized by the Van der Waals diameter $d_{VdW}$ ($d_{VdW}=560pm$ for Yb): 
\begin{equation}
    \bar{\ell} = \dfrac{k_b T}{\sqrt{2}\pi d_{VdW}^2} \dfrac{1}{p} \label{eqmfp} \\
\end{equation}
At $T=400^{\circ}C$, a corresponding partial Yb pressure of $p=10^{-3}mbar$, $\bar{\ell}= 67\ mm$ and 
$K_n=268 \gg 1$. When increasing the temperature, the mean free path will be reduced and when it becomes of the order of the length of the micro-tubes, the flow of atoms undergoes a transition from the previous 'transparent' regime ($\bar{\ell} \gg l_c$) into a so-called 'opaque' regime ($\bar{\ell} \lesssim l_c$), with a less favorable increase of the atomic flux\cite{Giordmaine}. 

\begin{figure}
\includegraphics[width=0.45\textwidth]{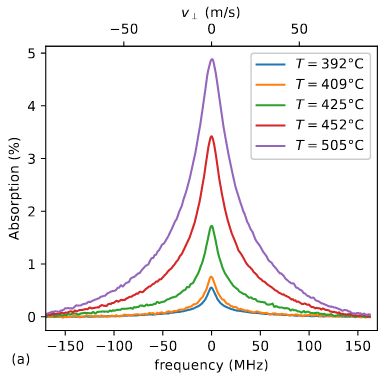}
\includegraphics[width=0.45\textwidth]{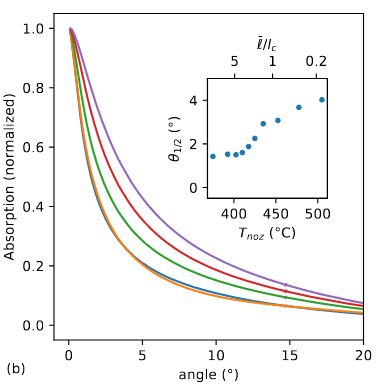}
\caption{\label{fig:SpectroCross1} (a) Absorption spectroscopy on the $^{174}Yb$ $^1S_0 \rightarrow ^3P_1$ line, orthogonal to the atomic beam, $11cm$ after the nozzle, for different nozzle temperatures (blue: $T=390^{\circ}C$, orange: $T=410^{\circ}C$, green: $T=425^{\circ}C$, red: $T=450^{\circ}C$, purple: $T=505^{\circ}C$). 
(b) Atomic beam angular distribution for the same temperatures. The inset  shows the half-angle half-maximum of the distribution as a function of the nozzle temperature and ratio of the mean free path over the micro-tube length.}
\end{figure}

In order to experimentally characterize our atomic beam, we make use of the window port of the UHV crosses before and after the pneumatic valve and the differential pumping tube. In Figure \ref{fig:SpectroCross1}, we show an absorption spectrum along the transverse direction of the atomic beam at 11cm after the oven output before our differential pumping tube. In order to avoid a convolution of $29MHz$ at the line at $^1S_0 \rightarrow ^1P_1$ line, we use for this particular measurement a narrow linewidth laser (Toptica DLC TA-SHG pro, with a linewidth below 100kHz) on the $^1S_0 \rightarrow ^3P_1$ at $\lambda_G=556nm$ with a natural linewidth of $2\pi 182kHz$. We have calibrated the relative frequency scan of the laser by molecular iodine spectroscopy\cite{Tanabe2022}. One clearly sees the increased absorption as we increase the temperature of the oven. For convenience, we convert frequencies (lower horizontal scale) to orthogonal velocities (upper horizontal scale) by $\delta(MHz) = v_{\perp}/\lambda_G$.
In Figure \ref{fig:SpectroCross1}(b), we plot the normalized absorption spectrum and convert the frequency (resp. velocity scale) into an angular scale (using a convolution by the longitudinal velocity distribution and the  transverse velocity distribution). We limit the plot to $22^{\circ}$ corresponding to the geometrical limits of our vacuum system. As shown in the inset, for temperatures below $425^{\circ}$C, the half width at half maximum (HWHM) of the angular distribution is of the order of $\theta_{1/2}=1.8^{\circ}$, close to $d/l$, even though twice as large as expected \cite{Giordmaine} from $\theta_{1/2}= 1.68d_i/2l_c=0.92^{\circ}$.
As we increase the temperature of the nozzle $T_{noz}$, we clearly see an angular broadening appearing above $T_{op}=425^{\circ}$C, corresponding to $\bar{\ell}(T_{op}) \simeq 2 l_c$, illustrating the transition from the 'transparent' to the 'opaque' molecular flow of our micro-tube array when the Knudsen number is of order of unity. In the upper scale of the inset of Figure \ref{fig:SpectroCross1}(b), we also plot the ratio of the estimated mean free path $\ell$ (eq. (\ref{eqmfp})) and the length of the micro-tubes. We stress  that the total flux of atoms in the wings of this distribution (to be integrated over azimuthal angle in contrast to this probe along one direction only) is important with $\sim98\%$ of the atoms leaving with an angle larger than $2^\circ$, even in the low temperature regime. Also, only $<2\%$ enter the cell with ballistic motion, relevant for residual collision losses with the trapped atoms, whereas one needs to take into account the isotopic abundance of $\sim 32\%$ for $^{174}Yb$ to estimate the flux relevant for the loading of the MOT of $^{174}Yb$.

\begin{figure}
\includegraphics[width=0.45\textwidth]{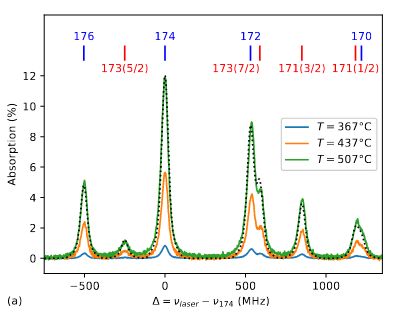}
\includegraphics[width=0.45\textwidth]{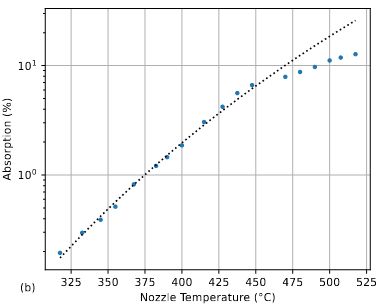}
\caption{\label{fig:SpectroScienceCell} (a) Absorption spectroscopy on the $^1S_0 \rightarrow ^1P_1$ line, at the science chamber center, orthogonal to the atomic beam, at different nozzle temperatures (blue: $T=367^{\circ}C$, orange: $T=437^{\circ}C$, green: $T=507^{\circ}C$). The 8 lines correspond to fermionic (red) and bosonic (blue) isotopes with their atomic number indicated in red and blue respectively. The black dashed curve is a fit by a Voigt profile, using a fixed Lorentz width of $2\pi 29MHz$ and free Gaussian width parameter. We have rescaled the resulting curve in order to match the maximum absorption of natural abundance of $^{174}Yb$ at $T=507^{\circ}C$, with corresponding smaller amplitudes for the other isotopes. (b) Maximum absorption for $^{174}Yb$ $^1S_0 \rightarrow ^1P_1$ as a function of the nozzle temperature. The black dotted line is fit with temperature (see text).} 
\end{figure}

With the flux of the oven well characterized, we now turn to the efficient use of this atomic beam for loading and trapping a large number of atoms in a MOT. As a last characterization before running our MOT, we turned to a spectroscopic study of the atomic beam in the cell chamber, 
where our differential pumping tubes inserted before the science cell only allow atoms with angles smaller than $2^{\circ}$ to enter. To characterize our atomic beam in the science cell, we have performed transverse spectroscopy over a large frequency range at $399nm$ (see Figure \ref{fig:SpectroScienceCell}(a)). Scanning the laser over more than 2 GHz allows us to observe the absorption lines of the various isotopes of Yb, with both bosonic and fermionic isotopes. We can identify 8 lines, which  we have fitted with a Voigt profile to take into account the convolution of the residual transverse Gaussian velocity distribution at the cell chamber and the natural linewidth of $2\pi 29MHz$ of the $^1S_0 \rightarrow ^1P_1$ transition. The free fit parameter of the Gaussian component yields a rms value of $k v_{perp}\simeq 2\pi 15 MHz$, corresponding to a transverse velocity of $v_{\perp} = 6m/s$, consistent with the geometrical selection given by our differential pumping section \cite{Wodey_2021}. 

Finally, in Figure  \ref{fig:SpectroScienceCell}(b), we show the peak absorption of the $^{174}Yb$ resonance as a function of the nozzle temperature. The observed increase is close to the one expected from eq. (\ref{EqPressure}) for temperatures up to $T=450^{\circ}$C, slightly above the temperature $T_{op}=425^{\circ}$C when crossing from the 'transparent' to the 'opaque' regime. As one can also see, the atomic flux still increases above $T=450^{\circ}$C. However as this corresponds only to a marginal increase, we operate our MOT mostly at $T=450^{\circ}$C. The black dotted line is a fit using $\log(Absorption)=-B/T[K]+C$, with the resulting fitting parameters of $B=5063$ and $C=1.6$. This is in agreement with the corresponding value of $B=8111$ from the literature (see eq. (\ref{EqPressure})).

\section{\label{MOTModel} Model of Magneto-Optical Trapping}

\subsubsection{Single atom regime}

Let us first recall the simple model for atom loading in a MOT, when only single atom physics is involved. In this case, the number of atoms trapped ($N$) is described by :
\begin{equation}
\label{eq:singleatomloading1} 
\dfrac{dN}{dt} = L-\alpha N \\
\end{equation}
where $L$ is the loading rate of the atoms in the MOT, which  can be extracted from the initial slope of the number of trapped atoms, and $\alpha$ correspond to the loss rates of atoms in the MOT.
Integration of eq. (\ref{eq:singleatomloading1}) yields 
\begin{equation}
\label{eq:singleatomloading2} 
N(t) = \dfrac{L}{\alpha} + \left[ N_0 - \dfrac{L}{\alpha} \right] e^{-\alpha t},
\end{equation}
with $N_0$ the number of atoms at $t=0$ (depending on whether one studies initial loading or decay).
The number of atoms that can be trapped is limited by the loss rate $\alpha$, which, within this model, also corresponds to the lifetime of the MOT, e.g. after switching off the loading. In order to study quantitatively the limits of our experiment, we distinguish three phenomena responsible for single atom losses:
\begin{equation}
\label{eq:singleatomloading3} 
\alpha = \alpha_{2,0} + \alpha_{bg} + \alpha_{atb}\\
\end{equation}
First, decay from $^1P_1$ to the meta-stable levels $^3P_0$ and $^3P_2$ can occur, which optically pumps atoms into long lived states. Atoms in $^3P_0$ are no longer trapped by our lasers and escape the MOT region. Atoms pumped into $^3P_2$ can be magnetically trapped allowing a kind of reservoir of cold atoms to be filled. In the experiments described in this paper, we have not implemented any repumping laser, and atoms in $^3P_2$ are thus also an effective loss term for the data presented here. We note that such losses via optical pumping are in principle very similar for Yb and Sr atoms. However, the branching ratio for optical pumping into the triplet manifold is several orders of magnitude lower for Yb \cite{Repumping} than for Sr \cite{Moriya_2018}, making the use of repumping lasers to increase the number of atoms less important for most Yb experiments.
We combine two optical pumping mechanisms into one loss term $\alpha_{2,0}$. This loss term will be proportional to the fraction $f$ of atoms excited into the $^1P_1$ state and thus depending on the laser intensity and detuning. For $I/I_{sat}\sim 0.1$ and $\Delta = -\Gamma$ we obtain an optical decay rate $A_{2,0}$ into the two long lived tripled states of \cite{Branching} $A_{2,0}\simeq 6.5s^{-1}$, which when taking into account an excited state fraction of $f=0.02$ yields a loss rate due to optical pumping of $\alpha_{2,0} \simeq 0.13$. We note that there is a third decay channel from $^1P_1$ to $^3P_1$ to occur. With a lifetime of $866 ns$ this state can decay sufficiently fast to the ground state $^1S_0$ before the atom can escape the MOT region. We do thus not include this decay channel in our estimation of losses by optical pumping.

A second and dominant loss mechanism in our experiment is collisions with hot atoms, either from the background pressure in the vacuum cell or from atoms in the atomic beam. We distinguish the collisional loss rate with residual atoms $\alpha_{bg}$ (background pressure) from the one corresponding to collisions with atoms in the atomic beam $\alpha_{atb}$. Collisions between cold trapped atoms (close to $v=0$) and hot atoms (with $v\sim 300m/s$) provides the initially cold atoms with a velocity that can be well above the capture velocity ($v_c \sim 40m/s$) of the MOT. Even though this loss mechanism also slightly depends on the laser parameters and magnetic field gradients\cite{Caires2004}, we neglect this dependence and approximate these losses by terms only dependent on the density of the hot atoms.

\subsubsection{Light-assisted collisions}

The single-atom regime discussed above needs to be revisited when interactions between atoms start to play a role. One can distinguish three different types of interactions. A first mechanism has been identified in the early years of cold atoms research and depends on the attenuation of the incident laser beams and multiple scattering of light inside the cloud of cold atoms. This mechanism is based on the 'long' range ($1/r$) driven dipole-dipole interactions and scales with the optical depth of the atomic cloud. A consequence of this interaction is an upper limit to  the spatial density of the atomic cloud rather than the number of trapped atoms\cite{Wieman90}. A second type of interaction is based on the  'short' range ($1/r^6$) Van der Waals interactions between atoms and depends on the spatial separation between atoms and thus on the spatial density of the atomic could. Indeed these interactions can be described as effective collisions with a corresponding scattering length which are important in evaporative cooling of atoms, as used to obtain quantum degenerate gases of cold atoms. These effects typically become relevant for atomic densities above $10^{12} at/cc$, a regime we do not enter in this paper. We thus neglect these interactions. An intermediate regime of driven dipole-dipole interaction depends on their ($1/r^3$) near field terms. This term becomes important at lower atomic densities than the Van der Waals interactions and has been identified as the mechanism for light-assisted collisions, leading to the radiative escape of cold atoms from a MOT. We will take into account this type of collective effects, as it limits the number of atoms we are able to trap in our experiment. A phenomenological approach to include this effect in the dynamical equation of the number of trapped atoms is the introduction of a loss term $\beta$ \cite{Weiner1999} added to eq. (\ref{eq:singleatomloading1}):
\begin{eqnarray}
\label{eq:beta}
\dfrac{dN}{dt} &=& L-\alpha N - \beta \int n^2(r) d^3r.
\end{eqnarray}
As one can see from this equation, the number of trapped atoms will no longer be a single exponential, neither during loading of the MOT (starting from N=0) nor during a decay from a MOT (when switching off the loading term $L$).
Assuming a spherical Gaussian density distribution of the cloud (with a peak density $n_0$ and rms radius $a$:  $n(r)=n_0 e^{-r^2/2a^2}$), the integration of this equation for the steady state number of atoms $N_{st}$ yields: 
\begin{eqnarray}
L - \alpha N_{st} - \dfrac{\beta}{(2a\sqrt{\pi})^3} N_{st}^2 &=& 0
\end{eqnarray}
 One can see from this equation that once limited by light-assisted collisions, the steady state number of trapped atoms stops to increase linearly with the loading rate $L$, and only scales as $\sqrt{L}$. In contrast to the Van der Waals interactions, these light-assisted collision terms depend on the occupation of the excited state of the atoms and thus on the intensity and detuning of the laser beams.

\begin{figure*}
\includegraphics[width=0.4\textwidth]{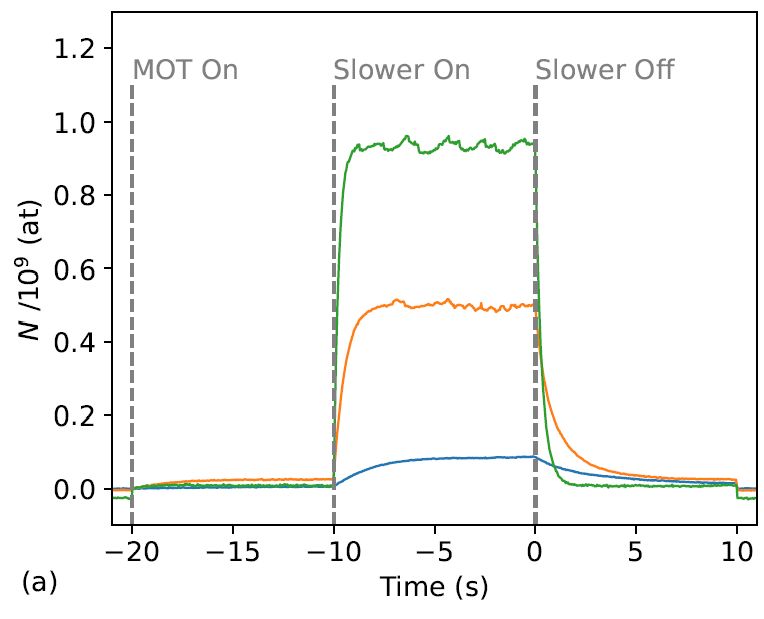}
\includegraphics[width=0.4\textwidth]{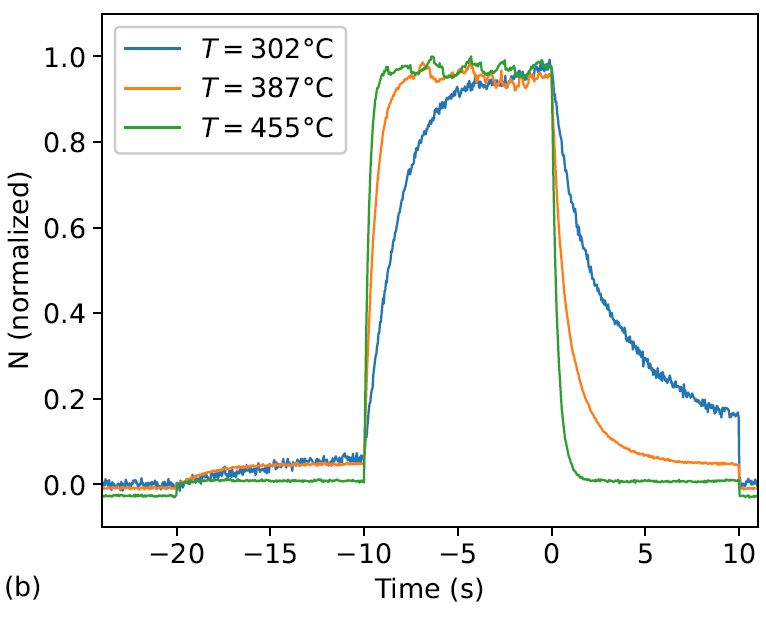}
\caption{\label{fig:Sequence} (a) MOT fluorescence during a typical MOT sequence for different nozzle temperatures (blue: $T=455^{\circ}C$, orange: $T=387^{\circ}C$, green: $T=302^{\circ}C$). Here only the temperature is changed, with all other parameters fixed:  $\Delta_{MOT}=-2\Gamma$, $B'_z=32G/cm$ and slowing beam power $P_{slow}=300mW$. For these experiments, the total saturation parameter of the atoms in the MOT is $s_{\Delta}=0.012$ with $19mW$ in each horizontal beam and $12mW$ in both vertical beams, corresponding to  $P_{tot}=100mW$. (b): The same data normalized to their maximum value to highlight the decay curves.}
\end{figure*}

\section{\label{Experimentalresults}Experimental results}

We now turn to the description of our experimental results, illustrating the different regimes discussed above.
We will study in particular the loading rate $L$, the number of trapped atoms in steady state $N_{st}$ and the losses $\alpha$ for various control parameters, such as slowing beam, laser intensity and detuning, magnetic field gradient and oven temperature. 

The capture velocity of a MOT depends on the optical forces required to cool and trap the atoms. For atoms without any Zeeman degeneracy in the ground state, the simple Doppler cooling and trapping model is relevant. The capture velocity of a MOT depends on several parameters, including laser intensity and detuning, magnetic field gradient and laser beam size. In many experiments, detuning and magnetic field gradients are easily changed and thus optimized to trap the maximum number of atoms. Laser power is often a more restricted resource, with some experiments resorting to three retro-reflected MOT beams. We stress that this is indeed a possibility to increase the number of trapped atoms, but only when multiple scattering effects are not important ingredients, i.e. when the number of trapped atoms is not approaching the $10^9$ regime. An even more important parameter is the size of the laser beams. This is typically very difficult to change and optical access to the MOT region is often an additional constraint. However, it has been shown that increasing the size of the laser beams allows for an impressive increase in the number of trapped atoms\cite{Camara2014}, provided the available laser power is sufficient to keep the peak laser intensity constant. In our present experiment, we have thus used a beam size with a waist of $w_0=22mm$, limited mainly by the size of our glass vacuum cell. 

In order to extract quantitative information of our MOT and the various mechanisms at play, we resort to the study of loading and decay curves of the atoms in the MOT. A typical sequence is shown in Figure \ref{fig:Sequence}(a), where we have plotted the number of trapped atoms for three values of the oven temperature. At $t=-20s$, we switch on the MOT laser beams (the magnetic field gradient is kept constant during the whole sequence). One can observe an increase in the detected fluorescence, calibrated and plotted as atom number for a fixed set of MOT laser parameters. At $t=-10s$, we switch on the additional slowing beam, resulting in an important increase in the number of trapped atoms. The slope after $t=-10s$ can be used to extract the loading rate of the MOT in the presence of the slowing beam. At $t=0$, we switch off the slowing beam, leading to a decay of the atom number towards the value with only the MOT lasers present. The decay curve can present different shapes (as high-lightened in the normalized curves shown in Figure \ref{fig:Sequence}(b), allowing us to identify a different regimes of atom losses.

\subsection{Role of laser detuning and magnetic field gradient}

\begin{figure*}
\includegraphics[width=0.4\textwidth]{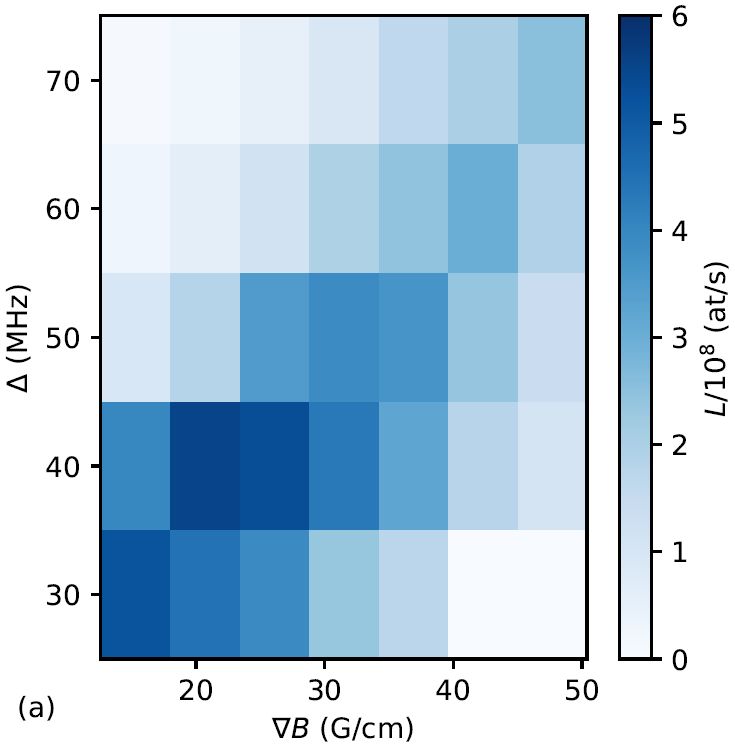}
\includegraphics[width=0.4\textwidth]{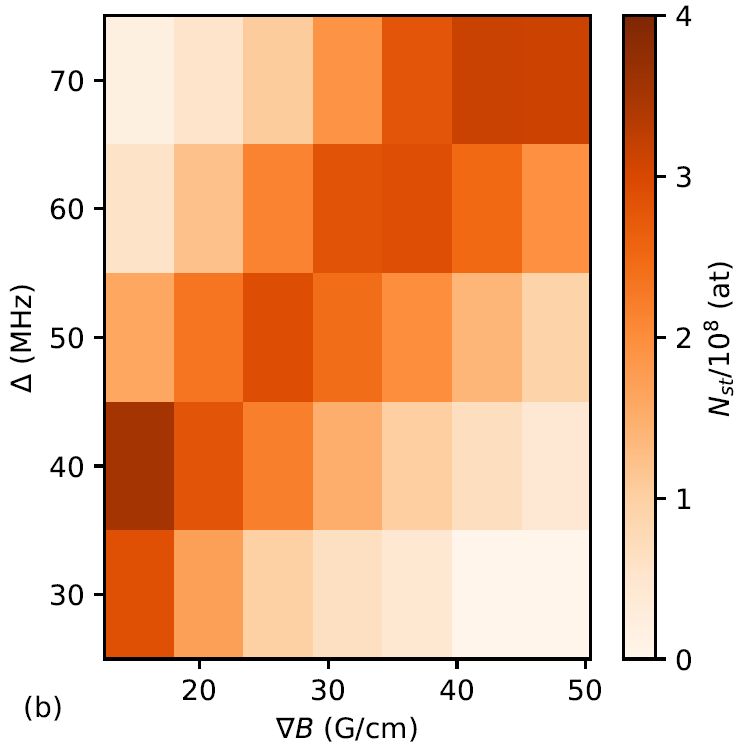}
\caption{\label{fig:vsBD} : (a) Loading coefficient $L(at/s)$ as a function of the MOT beam detuning and magnetic field gradient $\nabla B$ in the presence of a slowing beam. (b) Corresponding stationary number of trapped atoms $N_{st}$. Here the laser power is fixed to $P_{MOT}=62mW$ and $P_{slower}=200mW$.}
\end{figure*}

As a first standard optimisation procedure, we measured the loading rate and the number of trapped atoms as a function of both laser detuning and magnetic field gradient (see Figure \ref{fig:vsBD}) at constant laser power, 
with $P_{slower}=200mW$ for the slowing beam and a total of $P_{MOT}=62mW$ for the six MOT beams, corresponding to a total saturation parameter at $\Delta=-2\Gamma$ of $s_{\Delta}=0.0074$ and at an oven temperature of  $T \simeq 400^{\circ}C$. We note that we change the detuning of our MOT laser by changing the frequency of the Ti:Sapphire laser (before frequency doubling). We thus also change by the same frequency shift the frequency of our slowing beam. In Figure \ref{fig:vsBD} one observes that the optimal detuning depends on the magnetic field gradient along a line given by :
\begin{eqnarray}
\label{eq:detuninggradB}
\Delta \simeq 0.9 MHz/(G/cm).
\end{eqnarray}
which is of the same order of magnitude as the predicted scaling given by \cite{Dalibard_CdF2015}:
\begin{equation}
\label{eq:detuninggradB}
\Delta = \frac{\mu_0\nabla B}{\hbar}\frac{L}{2}\simeq 0.3MHz/(G/cm).
\end{equation}
when taking $L=2 w_0$ with our waist of $w_0=22mm$ ($\mu_0/\hbar=1.4MHz/G$). We note that the magnetic field gradient used to plot Figure  \ref{fig:vsBD} corresponds to the gradient along the magnetic field coils, which are placed vertically in our experiment and one should rather consider the magnetic field gradient along the atomic beam, which would then correspond to an increased slope of $1.8MHz/(G/cm)$, with an even larger discrepancy to the predicted scaling. 
The derivation of the above expression assumes however sufficient laser power to efficiently slow down the atoms during their initial crossing of the laser beams. The magnetic field gradient of the MOT plays the role of a small Zeeman slower and thus allows to increase the velocity capture range of the MOT compared to the velocity capture range of a Doppler cooling scheme in the absence of spatial confinement\cite{Dalibard_CdF2015}.

\subsection{Losses by light-assisted collisions}

\begin{figure}
\includegraphics[width=0.4\textwidth]{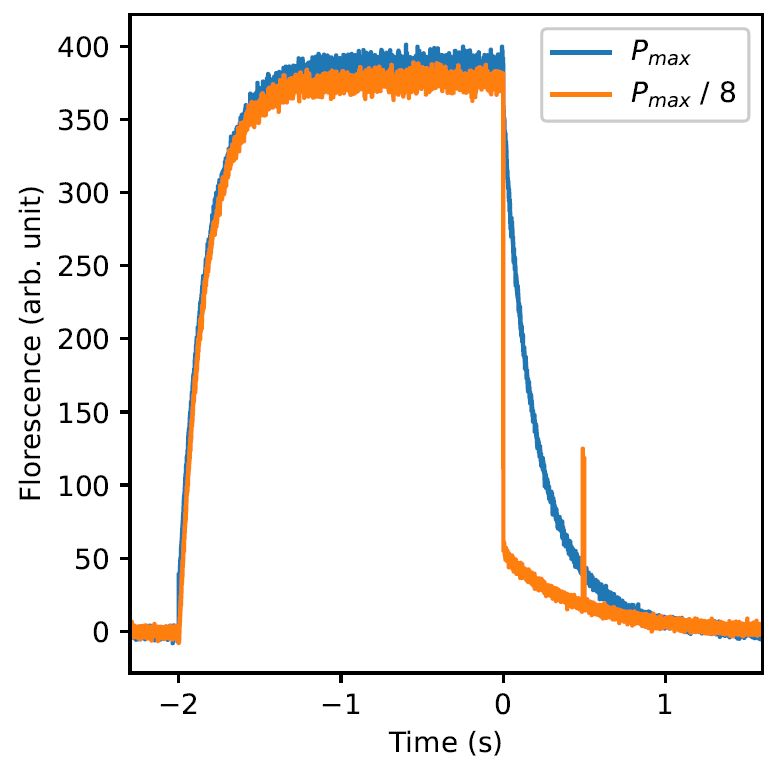}
\caption{\label{fig:beta} Fluorescence signal from the MOT with a change of MOT laser intensity by a factor of 8 at time $t=0s$ and slower turn off simultaneously. Losses from optical pumping are negligible. At $t=0.5s$,  we probe the remaining number of atoms using the same intensity. The increased fluorescence at $t=0.5s$ for $P_{max/8}$ (orange curve) compared to the value corresponding to $P_{max}$ (blue curve) is the signature of longer MOT lifetime at reduced laser intensity. The MOT parameters are $\Delta=-2\Gamma$, $P_{max}=170mW$ leading to $I=0.8 I_{sat}$ and $s_{\Delta}=0.042$ with $\nabla B = 18G/cm$. }
\end{figure}

\begin{figure}
\includegraphics[width=0.4\textwidth]{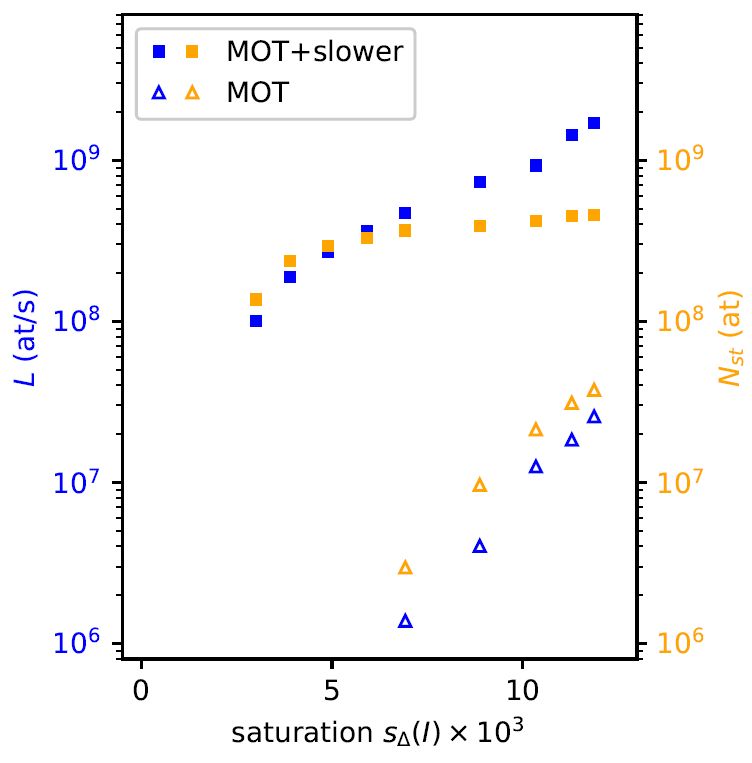}
\caption{\label{fig:Power} Loading rate $L$ (blue open triangles and full squares) and stationary atom number $N_{st}$ (green open triangles and full squares) as a function of the MOT power $P_{MOT}$ scale by $s_{\Delta}(I)$, with (full squares) and without (open triangles) slower beam. The other parameters are fixed, $P_{slower}=300mW$, $\Delta=-2\Gamma$, $\nabla B = 32G/cm$ and nozzle temperature $T=395^{\circ}C$. }
\end{figure}

We now turn to the study of light-assisted collisions, which turned out to be the main limitation for the maximum number of trapped atoms in our experiment. For this study, we changed the power of the incident laser beams for the MOT. Even though we have more than 2W of light available at $399nm$, in the final version of our optical setup for the blue light, we use a commercial optical bench with an optical input fiber. In order to prevent any damage to this fiber, we tried to limit the optical power in many of our experimental studies. In Figure \ref{fig:Power}, we show the loading rate and the steady state number of trapped atoms as we increase the laser power of the MOT beams from $P_{MOT}=25mW$ to $100mW$. For this series of experiments, we used $P_{slower}=300mW$, an oven temperature of  $T \simeq 400^{\circ}C$, a laser detuning of $\Delta=-2\Gamma$, and a magnetic field gradient $\nabla B = 32G/cm$. We plot the loading rate and atom number as a function of the saturation parameter computed at $\Delta=-2\Gamma$. First, we study the loading rate and atom number in the absence of any slowing beam (open triangles). The loading rate is extracted from the initial slope of the number of trapped atoms vs time when turning on the laser power, whereas the steady state number of atoms is obtained the value after reaching a plateau in the number of trapped atoms. The number of trapped atoms is obtained from a fluorescence signal, taking into account the atomic excitation (depending on laser intensity and detuning), the optical detection geometry and the detector efficiency.
One can see that both the loading rate and the trapped atom number increase with increasing the laser power. The ratio between the loading rate and the steady state atom number is constant, consistent with the single atom regime discussed above. This situation changes when we add a slowing beam, allowing to significantly increase both the loading rate and the number of trapped atoms. Whereas the loading rate continues to increase with increased laser power, we observe a saturation of the steady state atom number. We associate this limited atom number with the occurrence of light-assisted collisions, dominant at large atom numbers and laser intensity. 
As a qualitative signature of the role of light-assisted collisions, we compared the lifetime of the MOT at different laser intensities. Therefore we first load a MOT in a regime where we suspect light-assisted collisions to be present. We then partially switch off the loading of the MOT by switching off the slowing laser. The relaxation to the new steady state is still governed by eq. (\ref{eq:beta}) with an initial decay rate larger than the long time decay rate (blue curve on \ref{fig:beta}). In order to highlight the role of light-assisted collisions, we reduce the MOT laser intensity by a factor of 8 when switching off the slowing beam. This explains the fast drop of the fluorescence signal at t=0 (orange curve on Figure \ref{fig:beta}). As we reduce the light-assisted collisions by reducing the laser intensity, we expect the lifetime of the MOT to be increased. In order to highlight this effect, we compare the number of atoms with a delay of $0.5s$ after swichting off the slowing beam. In order to compare the atom number under the same probing conditions, we change for a short time the power of the MOT beams to the full power $P_{max}$. This allows for a direct comparison of the fluorescence and corresponding atom number for different decay regimes: high and low intensity. As one can observe in \ref{fig:beta}, the number of trapped atoms decayed less when the MOT beam where reduced by a factor of 8 compared to the decay at constant MOT laser power. This is a direct qualitative proof of the presence of light-assisted collisions in this experiment. 

\subsection{Losses by Optical pumping}

\begin{figure}
\includegraphics[width=0.4\textwidth]{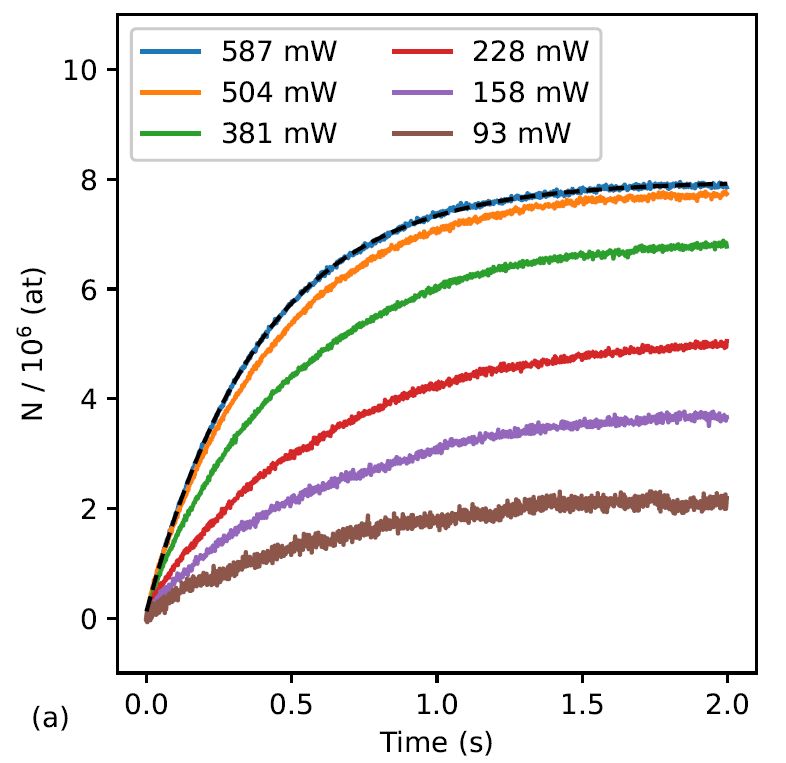}
\includegraphics[width=0.4\textwidth]{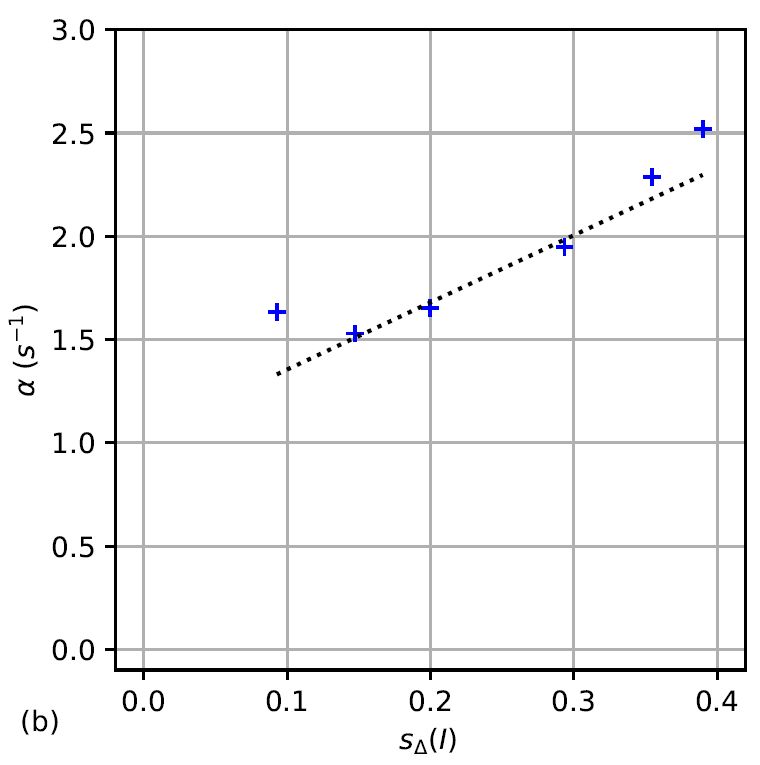}
\caption{\label{fig:LvsPB05_courbes} (a) MOT loading curve for different laser power. The maximum $P_{max}=587mW$ power corresponds to $ I_{tot}=2.8I_{sat}$. The magnetic field gradient is fixed at a very low value $\nabla B = 3.1G/cm$ and detuning is optimized with the atom number at $P_{max}$ to $\Delta=-1.4\Gamma$. Fitting these curves by an exponential (black dashed) allows to extract the loading and loss rates. (b) Extracted loss rate as a function of $s_{\Delta}(I)$. The dotted line corresponds to ab initio computed slope  $\alpha_{2,0}=6.5 s_{\Delta}(I)/2 s^{-1}$,  shifted by  $1s^{-1}$ for rest-gas collisions.}
\end{figure}

After having identified the regime of light-assisted collisions, we now turn to the study of losses induced by optical pumping \cite{PowerLoss,Repumping} as described by $\alpha_{2,0}$ in eq. (\ref{eq:singleatomloading2}). This regime can be studied by tuning our MOT parameters to obtain very low spatial densities. In particular, this can be achieved by reducing the magnetic field gradient, resulting in a weaker confinement force. 
At this point it is important to stress that MOTs based on pure Doppler forces are much far sensitive to intensity imbalances of counter-propagating beams and local imperfections than in the presence of Sisyphus cooling\cite{Lett1989}. A reduced magnetic field gradient thus imposes a more stringent control of beam qualities and intensity balance \cite{Kawasaki_2015}. We have been able to operate a MOT at magnetic field gradients of the order of $\nabla B\sim 2G/cm$, where the MOT size reaches the order of the laser beam size. For the study of losses by optical pumping, we have used a magnetic field gradient of $\nabla B=3.1G/cm$, a laser detuning of $\Delta=-1.4\Gamma$, and total power in the MOT beams of $P_{MOT}\sim 600mW$ (which has been realized without the fibered commercial optical bench, allowing for higher laser power to be used). Furthermore, to work with a very small number of trapped atoms, we also do not use a slowing beam for this study. As shown in Figure \ref{fig:LvsPB05_courbes}(a), we observe an exponential loading of the MOT with less than $10^7$ atoms trapped at the largest power used in this study. From this loading curve, we can apply an exponential fit (as shown in the black dashed line in Figure \ref{fig:LvsPB05_courbes}(a)) and extract the loading rate and the loss rate using eq. (\ref{eq:singleatomloading2}).  In Figure \ref{fig:LvsPB05_courbes}(b), we plot this loss rate as a function of the saturation intensity $s_\Delta$ computed at the corresponding detuning. We observe an increase in this loss rate, which we now attribute to the optical pumping, as we are safely in a regime without light-assisted collisions. The dotted line in  Figure \ref{fig:LvsPB05_courbes}(b) corresponds to the ab initio computed dependence of  $\alpha_{2,0}=6.5 s_{\Delta}(I)/2 s^{-1}$, shifted by an offset value of $1 s^{-1}$ attributed to collisions with hot atoms. The good agreement with this computed slope and our experimental results supports our interpretation of this curve to be explained by optical pumping. 

\subsection{Losses by collisions with hot atoms}

After these studies of the losses due to light-assisted collisions and optical pumping, we now turn to the last loss term in eq. (\ref{eq:singleatomloading2}), namely collisions of trapped atoms with hot atoms from either the background pressure in our science cell or from atoms in the atomic beam. One convenient method to study these losses is to increase the oven temperature, leading to an increase in the atomic beam flux and also, despite the differential pumping scheme, in the residual background pressure in our science cell. We use again the similar methodology as shown in Figure \ref{fig:Sequence} to extract the loading and loss rate of our MOT. For this study, we keep all other parameters of the MOT fixed at $P_{MOT}=100mW$ (with $19mW$ in each of the four horizontal beams and $12mW$ in each vertical beam), a beam waist $w_0=22mm$, a detuning of $\Delta = 60MHz = -2 \Gamma$ corresponding to $s_{\Delta} = 0.012$, a magnetic field gradient of $\nabla B_z=32G/cm$ and with a slowing beam power of $P_{slow}=300mW$.

\begin{figure}
\includegraphics[width=0.39\textwidth]{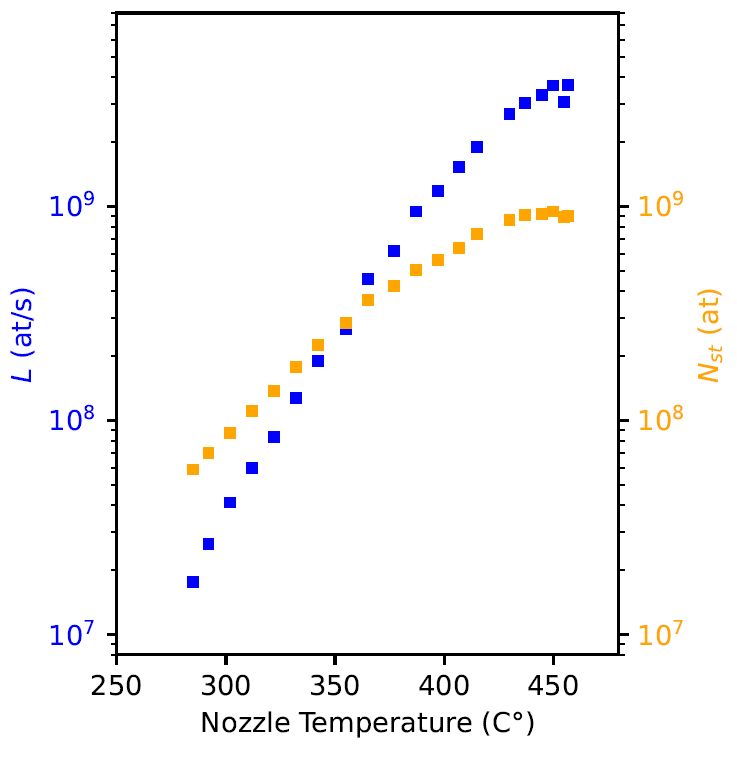}
\caption{\label{fig:LN} Loading rate $L$ (blue squares) and stationary atom number $N_{st}$ (orange squares) versus temperature with and without the slowing beam.}
\end{figure}

\begin{figure}
\includegraphics[width=0.49\textwidth]{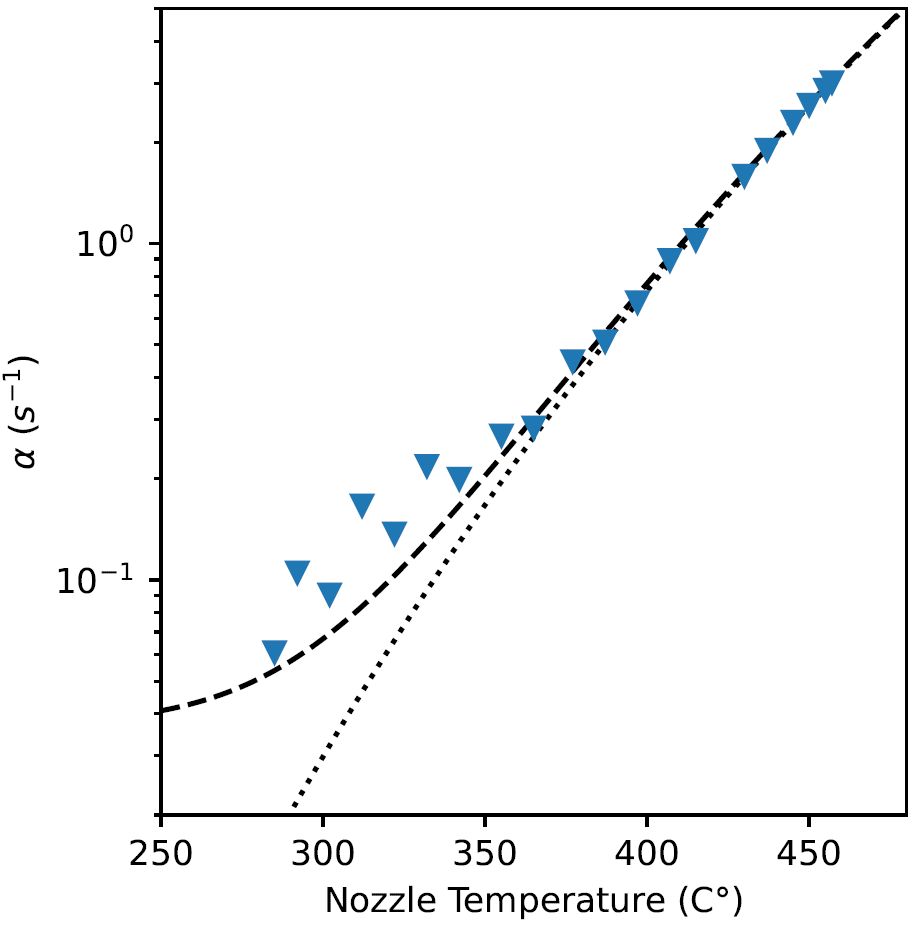}
\caption{\label{fig:pertesgas} Collisional loss rate as a function of the nozzle temperature. The dotted line if a fit at high temperatures corresponding to $\log(absorption)=-5354/T[K]+7.816$. Optical pumping loss are added to better describe losses at low temperatures (dashed line).}
\end{figure}

In Figure \ref{fig:LN} we show the corresponding loading rate $L$ and steady state atom number $N_{st}$ as we increase the nozzle temperature from $T=285^\circ C$ to $T=480^\circ C$. In the presence of the slowing beam, the results are similar to the results shown in Figure \ref{fig:Power}, where we varied the MOT power: the loading rate $L$ continues to increase up to the largest nozzle temperatures studied, whereas the steady state atom number $N_{st}$ saturates at about $N_{st}\simeq 10^9$, which we attribute to the occurence of light-assisted collisions. In this regime, where light-assisted collisions play an important role, we can nevertheless access the loss rate due to collisions with hot atoms by investigating the late time decay after switching off the slowing beam. Indeed, after an initial decay with important light-assisted collisions, once the number of atoms drops below $\sim 10^8$, we recover a single exponential decay from which we extract a loss term $\alpha$. In Figure\ref{fig:pertesgas} we show this loss term as a function of the nozzle temperature. Assuming this loss term to be proportional to the vapour pressure of Yb in the oven, we fit this curve using the scaling low given by 
$\log(pressure)=-A2/T[K]+B2$, yielding $A2=5354$ and $B2=7.816$ (dotted line in Figure\ref{fig:pertesgas}) The relevant value of $A2$ is close to the one we extract from the atomic beam spectroscopy (see Figure \ref{fig:SpectroScienceCell}). This good agreement supports our interpretation that these late time losses are determined by collisions with hot atoms. As shown in the dashed line in Figure \ref{fig:SpectroScienceCell}, we can further improve our fitting by adding a contribution associated to optical pumping of  $\alpha_{0,2} \simeq 0.04s^{-1}$.

As a final note, one can distinguish two contributions of losses by hot atoms, one due to collisions with background gas in the science cell (which increases when operating the atomic beam) and those between cold atoms and atoms in the atomic beam. In order to discriminate between these two contributions, we have added a mechanical shutter in the atomic beam, aiming at only cutting out atoms directly passing through the MOT. This shutter also reduces the loading rate, but we initially hoped that the reduction in direct collisions with the atomic beam would be more important (in principle up to 100 $\%$) than the partial reduction of the loading rate. We observed a reduction of the loading rate when turning in this mechanical shutter, but we did not observe an increase in atom number in the corresponding loss rate, hinting at a situation where we are limited by collisions with background gas and light-assisted collisions.

\subsection{MOT size and self-oscillating clouds}

\begin{figure}
\includegraphics[width=0.48\textwidth]{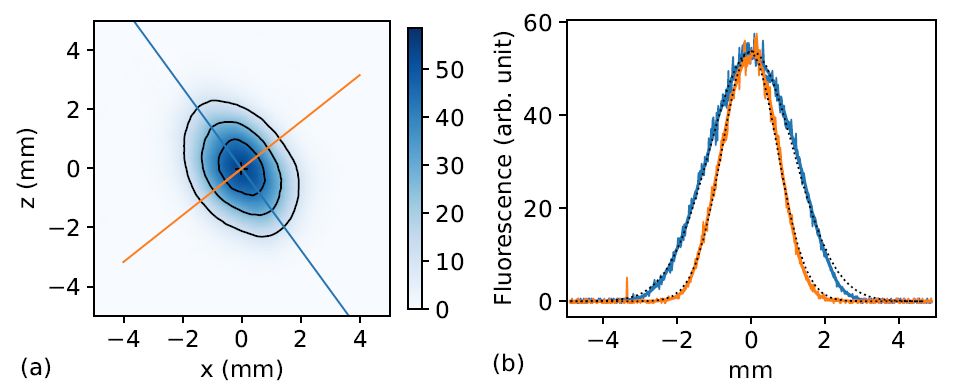}
\includegraphics[width=0.48\textwidth]{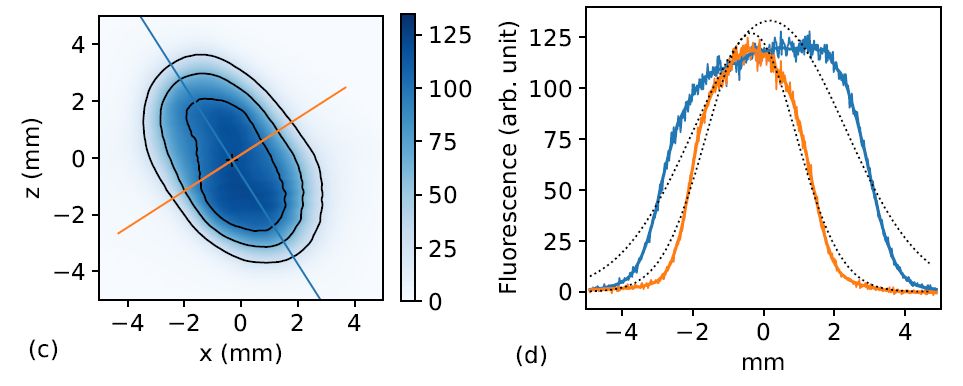}
\caption{\label{fig:PhotoT} Images of MOT at a temperature of (a) $T=320^{\circ}C$ and (c) $T=450^{\circ}C$, with contour lines at $20\%$, $50\%$ and $80\%$ of the maximum values. Profiles across the long axis (in blue) and short axis (in orange) are shown in (b) for $T=320^{\circ}C$ and  in (d) for $T=450^{\circ}C$. The dotted lines in (b,d) correspond to Gaussian fits of these profiles.}
\end{figure}

\begin{figure}
\includegraphics[width=0.49\textwidth]{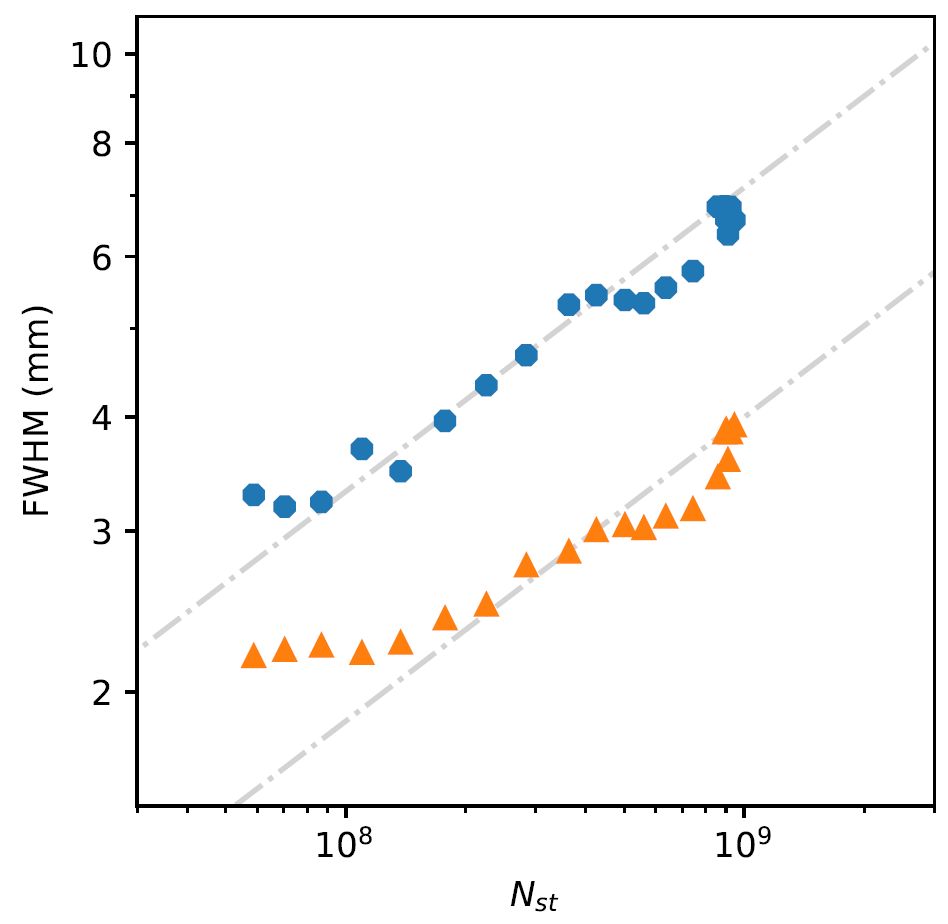}
\caption{\label{fig:fwhm} Full Width Half Maximum of MOT as a function of the atom number $N_{st}$ for the long (blue) and short (orange) axis of the MOT for the same data as in Figure \ref{fig:PhotoT}. The grey dotted-dashed lines indicate the $\sqrt[3]{N_{st}}$ scaling.}
\end{figure}

When studying large clouds of cold atoms, long range driven dipole-dipole interactions also become important. Even though they do not lead directly to a loss of atom number, one consequence is the increase of the size of the MOT due to the repulsive forces of multiple scattering \cite{Wieman90} dominating over the compressive forces due to attenuation \cite{Dalibard1988}. As shown in \cite{Wieman90, Camara2014} such repulsive forces leads to a maximum spatial density inside the MOT and thus to an increase of the size of the cloud as more atoms are loaded in the MOT. In Figure \ref{fig:PhotoT} we show fluorescence images and cuts of our MOT, at low oven  temperature (Figure \ref{fig:PhotoT}(a) and (b)) where the atom number is not yet sufficient to produce noticeable repulsive forces and Gaussian fits of the profiles are accurate, and higher oven temperature  (Figure \ref{fig:PhotoT}(c) and (d)) where one clearly observes a flatter center of the MOT. In Figure \ref{fig:fwhm}, we show how the size of the MOT increases with the atom number and find a consistent with the expected scaling of $\sqrt[3]{N_{st}}.$

\begin{figure}
\includegraphics[width=0.48\textwidth]{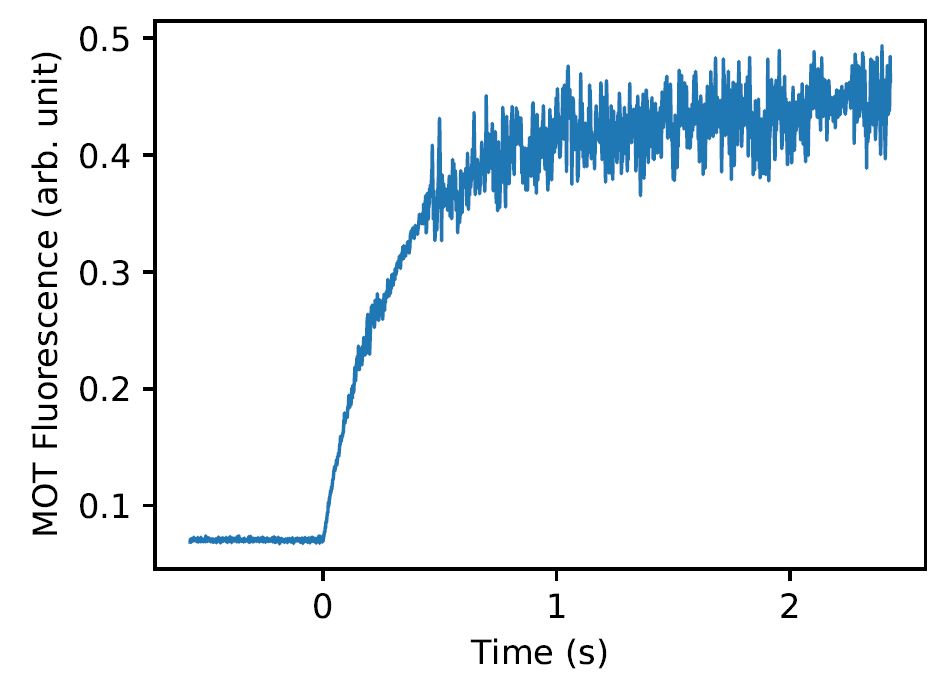}
\caption{\label{fig:selfosc} Self-Oscillation MOT Fluorescence, close to resonance $\Delta=-\Gamma/2$, $P_{MOT}=62mW$, $P_{slower}=200mW$, $\nabla B = 25G/cm$ and nozzle temperature $T=410^{\circ}C$.}
\end{figure}

Finally, as studied in \cite{Labeyrie06}, for even larger atom number and when tuning the laser closer to the atomic resonance, self-oscillating cloud can be observed. We have also been able to observe the transition to such self-oscillating clouds as shown in Figure \ref{fig:selfosc} where one can observe an increase of the fluctuations in the fluorescence signal above a threshold of trapped atoms, similar to what has been reported in \cite{Labeyrie06, Gaudesius2022}. We note that here we have used Yb atoms better described by Doppler forces than the previous experiments using Rb atoms and a more quantitative comparison with numerical models \cite{Gaudesius2022} should be possible.

\section{\label{Conclusion}Conclusion}

In conclusion, we have presented an experimental setup, able to laser cool and trap up to $N=10^9$ Ytterbium atoms, without resorting to a Zeeman slower. We have precisely characterized our atomic beam and the MOT on the $^1S_0 \rightarrow ^1P_1$ transition at $399nm$. We have in particular characterized the loading rate and various contribution to atomic losses. This quantitative understanding of the limits of our setup allows us to further enhance the trapped atom number in future upgrades of the experiment. One solution to reduce the dominant loss term of light-assisted collisions is to implement a central dark area of the MOT, similar to dark spot techniques implemented for alkali atoms\cite{Ketterle1993}. If the atoms in the center of the MOT are exposed to lower intensities while keeping the intensity large in the outer regions of the MOT, one can reduce light-assisted collisions in the MOT and keep an important loading rate with a large capture velocity. This technique also allows to reduce the atomic repulsion leading to a larger spatial density of the atomic cloud. A second approach to reduce the light assisted collisions is to implement a MOT on the intercombination line $^1S_0 \rightarrow ^3P_1$ at $556nm$, where light-assisted collision on this weaker line are well reduced. Also, using a shifted MOT allows to avoid direct collisions with hot atoms from the atomic beam. If needed, a new ion getter pump and/or a better differential pumping will allow to reduce the collisions with hot background atoms. We will also consider increasing the loading rate by both transverse cooling of the atomic beam and by using multiple frequencies in the slowing beam. Such an increased loading rate is of particular interest in studies of light-matter interactions, where the relevant timescales are typically well below $1ms$, making the MOT preparation time the dominant limitation to the duty cycle. 

\section{\label{Acknowledgements}Acknowledgements}
We thank G. Labeyrie, M. Hugbart and W. Guerin for fruitful discussions in the design and during the implementation of this experiment. We also thank A. Apoorva and A. Glicenstein for critical reading of the manuscript. This work was performed in the framework of the European project ANDLICA, ERC Advanced grant No. 832219.

\section{\label{Authordeclarations}Author declarations}
Conflict of Interest

The authors have no conflicts to disclose.

\section{\label{AuthorContributions}Author Contributions}

Hector Lettelier: design, implementation, spectroscopy, data acquisition, analysis and writing;
Álvaro Mitchell Galvão de Melo: laser control, spectroscopy and data acquisition;
Anaïs Dorne: participation in spectroscopy and initial data acquisition;
Robin Kaiser: funding acquisition (lead); supervision (lead); writing, review and editing.

\section{\label{Dataavailability}Data availability}
The data that support the findings of this study are available from the corresponding author upon reasonable request.

\section{\label{Acknowledgements}References}
\nocite{*}
\bibliography{BlueMOT.bib}% Produces the bibliography via BibTeX.

\end{document}